# Surface Induced Magnetism of CdSe Quantum Dots: A DFT Study

G. Kurian, D. Subedi, A. Aniagboso, M. D. Mochena*
Department of Physics, Florida A&M University, Tallahassee, Florida

*Corresponding author: mogus.mochena@famu.edu

Abstract

Surface induced magnetism of quantum dots (QDs) is not well understood and has been a subject of considerable controversy because of the complexity of surface chemistry of QD facets coupled with the many degrees of freedom inherently involved in magnetism. We performed spin-polarized colinear DFT calculations to study charge and spin densities to determine surface reconstructions in bare as well as passivated QDs in controlled way. We find that tri-coordinated surface atoms relax contributing no dangling bonds and bi-coordinated atoms dimerize, reducing the population of the dangling bonds affecting the induced magnetism. As a result, bi-coordinated rather than tri-coordinated surface atoms are responsible for spin-polarized surface states. In light of recent advanced experimental tools, we predict dangling bonds that remain, despite passivation, on low – index facets with bi-coordinated atoms could result in tunable surface magnetism. Our results also shed light on the contradicting experimental results in the literature.

## 1. Introduction.

Colloidal synthesis of QDs has become the most widely used technique of synthesis due to its facile nature and cost effectiveness. The surface chemistry of such a colloidally synthesized QD is complex. It consists of facets of an inorganic nanocrystal capped by organic ligands which arrest the growth of QDs at desired size and prevent them from agglomerating. Moreover, the interface between the inorganic nanocrystal and the capping ligands creates an additional layer of complexity.

Recently, nuclear magnetic resonance spectroscopy and diffusometry have been used to quantify the populations and kinetics of ligands binding to Pb QDs, highlighting the complexity of ligand binding mechanisms and possibility of tuning QD surface properties.[1] On computational front, ligand dynamics on the surface of CdSe nanocrystals has also been recently studied with classical molecular dynamics to locate the positions, binding modes and mobilities of carboxylate ligands on the different facets of CdSe nanocrystals.[2]

Over the past decades, there has been a great deal of interest in introducing magnetism in nonmagnetic quantum dots for spintronic applications through doping by magnetic elements.[7] Doping of QDs with magnetic elements colloidally, however, has been challenging mainly due to what is known as self-purification.[3,4] The dopants migrate to the surface, as a result, uniform distribution of the dopants in the volume of the QDs was difficult to realize. There have been, however, reports of successes in raising the solubility of the magnetic dopants. Nevertheless, such doping chemically is found out to affect optical and other native properties of the QDs. As a result, an alternative approach of inducing magnetism to chemical doping, that retains the native properties of QDs, through surface engineering was attempted over a decade ago by three groups with contradicting results.[5,6,7] Other than that, experimental efforts to study surface magnetism of QDs have been limited. However, recent developments in advanced experimental tools[1] coupled with first principles computation[2] could pave the way for such an alternative mechanism to introduce magnetism into QDs.

In this work, we looked into the possibility of tuning the surface of CdSe colloidal QDs to introduce surface magnetism. CdSe QD is chosen because cadmium chalcogenide nanocrystals result in a small group of low – index facets during growth.[8] Furthermore, these low – index facets can be classified into polar and non – polar ones. Preliminary results show that it is possible to control formation of the polar or non – polar facets for cadmium chalcogenide nanocrystals.[8] Such a prospect presents the possibility of a controlled facet to study surface chemistry and likelihood of tuning for desired surface magnetism. In the following sections, we will briefly present a summary of the contradicting experimental results, then our computational method and results, and finally our conclusion.

## 2. Summary of Experimental Results.

There are no recent experimental or theoretical works looking into the possibility of surface ferromagnetism to the best of our knowledge. Therefore, we refer to the experimental groups that had contradicting results on the nature of surface magnetism of QDs more than a decade ago.

Neeleshwar et al. studied CdSe QDs topped with trioctylphosphine oxide (TOPO) and found the QDs to be paramagnetic and attributed the observed magnetization to dangling bonds on the surface. They calculated the concentration of the dangling bonds to be size dependent: less than 2000 ppm for $d$=2.8 nm, less than 1000 ppm for d=4.1, and less than 300 ppm for d=5.6 nm with magnetic moment of $1.73\mu_B$ per dangling bond.[5]

Seehra et al. studied CdSe QDs of various sizes from d = 1.8 to 7.0 nm capped with TOPO and found room temperature ferromagnetism. Their SQUID measurements taken in temperature range from 0 to 370 K showed that the magnitude of remanence and saturation magnetization to be inversely proportional to the size of the QDs, a signature for ferromagnetism. The observed size dependent ferromagnetism was attributed to charge transfer from the 4d orbital of Cd to the O in TOPO based on the XANES and EXAFS measurements.[6]

Meulenberg et al. studied CdSe QDs topped with TOPO and hexadecylamine and found the QDs to be paramagnetic, contradicting Seehra et al.[7] They attributed the paramagnetism to the transfer of an electron from the 4d orbital of the surface Cd atoms to the empty $\pi^*$ orbital of the P = O bond in TOPO. Thus, the induced magnetism is due to ligands rather than to dangling bonds. Furthermore, they reported a magnetic moment of ~0.01 $\mu_B$ per Cd in agreement with Seehra et al. contradicting Neeleshwar et al.

It is obvious from the above contradicting experimental results that the surface chemistry of ligands and dangling bonds is not well understood. In colloidal synthesis, there could remain a fraction of uncapped surface atoms resulting in dangling bonds that could affect the properties of the QDs as in CdSe QDs[6,7,9]. Experimental results of XPS[10] and NMR[11] studies showed that when the CdSe QD is passivated with TOPO, the Se atoms are predominantly not passivated. XPS[10] and EXAFS[12] studies on TOPO passivated CdSe QDs showed that TOPO binds only at the Cd site and that the surface coverage on the Cd sites varies from about 60% for d = 2 nm to about 35% for d > 3 nm.[6] These results are consistent with those of Neeleshwar et al. and Seehra et al. On the other hand, the conclusion by Muelenberg et al. rests on the majority of the uncapped Se atoms remaining dangling bonds. Therefore, their concentration of dangling bonds is much higher than that reported by Neeleshwar et al.

In this work, we performed density functional theory (DFT) calculations to identify the sources of surface magnetism in CdSe QDs to resolve the experimental discrepancies and shed light on the surface chemistry involved. Although the QDs synthesized by Neelashwar et al.[5], Seehra et al.,[6] and Muelenberg et al.[7] are larger than the one used in our study, the size in this study is large enough as it includes the relevant bi-coordinated and tri-coordinated Cd and Se atoms across the different facets of the surface. A stoichiometric ratio of 1:1 between Cd and Se was set in accordance with the ratio in QDs used in experiments.[6] The QDs are capped with pseudo hydrogen atoms completely and then uncapped as needed to study the effect of dangling bands or realistic ligands that could be computationally handled. We present in the following sections the computational details first followed by results and conclusion.

## 3. Computational Details

We performed DFT calculations as implemented in Vienna ab initio simulation package (VASP). VASP is a code for periodic structures, however, it can be applied to finite structures such as QDs by surrounding them with enough vacuum space to avoid interactions between the QDs in adjacent supercells. First, we constructed the CdSe QD by cutting out a spherical quantum dot of approximately 2 nm, $Cd_{90}Se_{90}$, from bulk wurtzite structure. In order to simulate the ligand terminated QD in colloidal environment, we passivated the surface atoms with pseudo-hydrogens at 100% coverage initially. The dangling bonds were then created selectively as desired by uncapping the pseudo-hydrogens. Similarly, to see the effects of ligands, we uncapped a pseudo-hydrogen and substituted it with a short ligand analogue. The passivated or partially passivated QD was surrounded with a vacuum of 12 Å. The structure was then optimized to determine the ground state energy. The core electrons were replaced by projector augmented wave (PAW) pseudopotentials supplied by VASP. The 4d electrons in Cd were treated as valence electrons on similar footing as 5s electrons. Similarly, the 4p electrons of Se were treated as valence electrons as the 4s electrons. The exchange correlation functional was approximated by the Pedrew and Wang (PW91) generalized gradient approximation (GGA).[13] The Kohn-Sham equation is then solved self-consistently for a given configuration of ions in a plane wave basis with cut off energy of 300 eV until energy convergence of $10^{-4}$ eV was reached. The ionic configuration was updated with conjugant gradient scheme until a convergence criterion of 0.01 eV/Å for forces on the ions was reached. The Brillouin zone integration was performed at gamma point only for such a large supercell.

## 4. Results and Discussion

### 4.1. Facets and surface coordinations (CNs) of wurtzite CdSe

To facilitate the discussions of the different surfaces corresponding to the different facets of the QD, we refer first to bulk unit cells shown in Figures 1a) and 1b). At room temperature, CdSe crystallizes in wurtzite (WZ) phase. The WZ phase of a compound is multifaceted due to its unique c axis.[14] The relevant facets of the WZ phase of CdSe QD to this work are shown in Figures 1c) to 1f). The (0001) and the $(000\bar{1})$ surfaces are straightforward and terminate with single dangling bonds and CN=3 as seen in Figure 1c). The facets parallel to the c axis are either $\{1\bar{1}00\}$ or $\{11\bar{2}0\}$. The $\{1\bar{1}00\}$ could have either CN = 2 or CN = 3 as shown in Figures 1d) and 1e). Atoms in the planes of the $[1\bar{1}00]$ direction of Figure 1d) are bonded to one atom in the same plane and two atoms underneath the plane, thus any atom on that plane has CN = 3. If the top surface layer of Figure 1d) is removed, then each atom in the layer below will lose two bonds and thus their coordination number reduces to 2, resulting in Figure 1e). This shows where the cut of the QD out of bulk is made or the growth during synthesis is stopped matters in $[1\bar{1}00]$ direction.

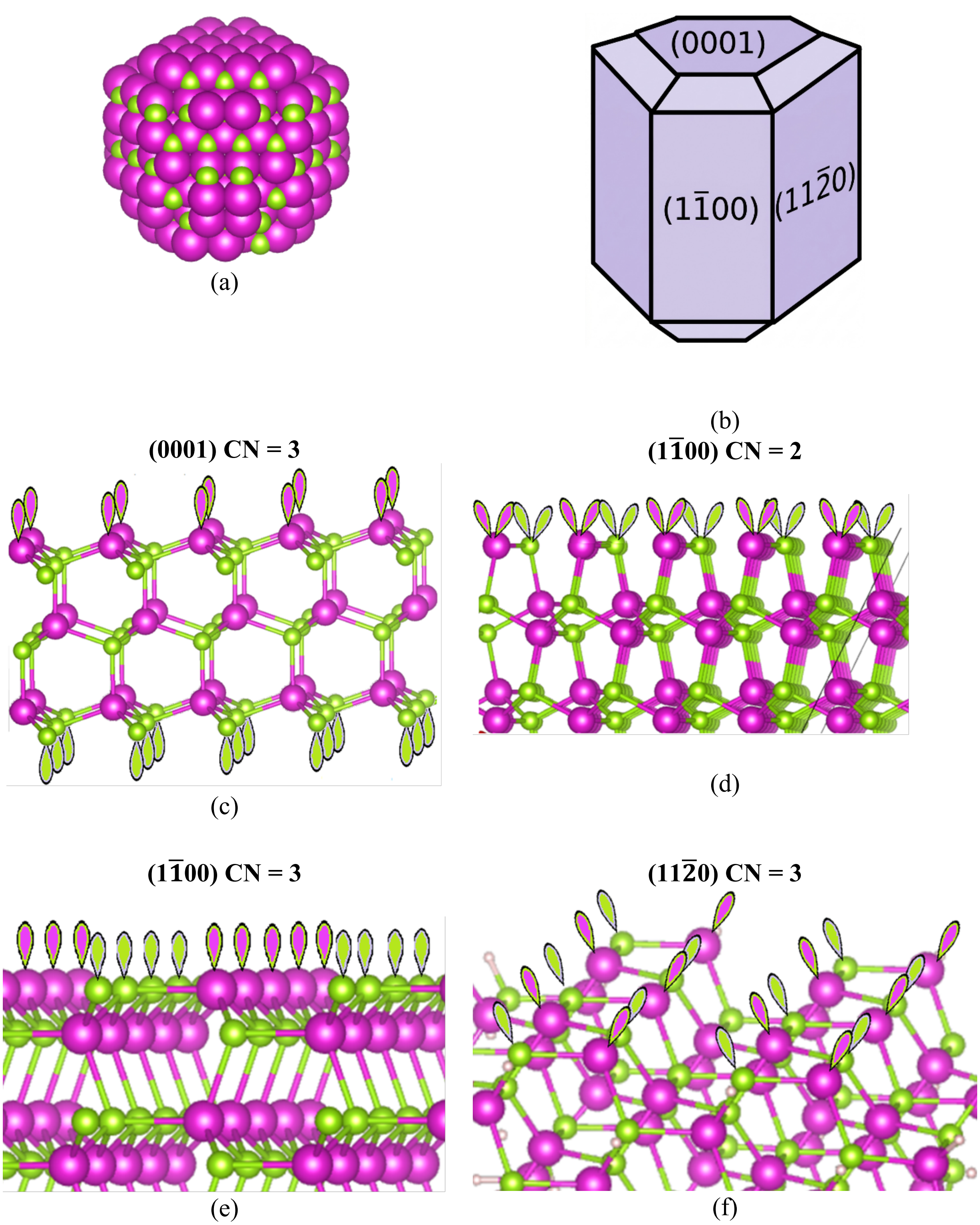

**Figure 1.** Color scheme: Cd pink, Se green, dangling bonds attached to Cd pink, and those attached to Se green**.** a) CdSe unit cell, b) lattice facets on the unit cell, c) the (0001) plane with CN=3 and single bonds, the bottom surface $(000\bar{1})$ has also the CN=3 with single dangling bonds but not shown, d) the $(1\bar{1}00)$ plane with CN = 2 and two dangling bonds, e) the $(1\bar{1}00)$ plane with CN= 3 and single dangling bonds, and f) the $(11\bar{2}0)$ plane with CN = 3 and single dangling bonds.

The $\{11\bar{2}0\}$ facets have CN = 3, with alternating Se and Cd atoms on the surface, each with a single dangling bond as shown in Figure 1f). The atoms in the $[11\bar{2}0]$ direction have the same number of bonds in each layer as seen in Figure 1f), therefore, removal of any surface layer only reduces the CN of the atoms in the surface under by only 1. By contrast, removal of the top layer in Figure 1e) results in a loss of three bonds for the atoms in the layer below, resulting in an atom with CN = 1. Such a configuration is energetically unstable, therefore, will evolve into a more stable one during synthesis and will not be considered.

### 4.2. Choice of input spin configurations

A priori, the location of the dangling bonds on the bare surfaces is unknown as relaxations or reconstructions could take place. Nevertheless, the likelihood of finding dangling bonds is high when all the surface atoms are uncapped. In VASP, one can let the code determine the locations; the initial magnetic moments are assigned large values so that the code optimizes them to actual values when the geometric optimization is completed. Therefore, initially, we assigned sizeable magnetic moments to all the atoms, coupled them ferromagnetically or antiferromagnetically as given in Table 1. One could argue if the code determines the final spin configuration why is it necessary to choose two different initial spin configurations. This is because the spin-orbit interaction necessary for changing spin orientation is not included in the colinear calculations performed in this work.

**Table 1.** Intital spin configurations. The total energies and magnetic moments of the optimized structures. The energies are referenced to configuration 1.

| No. | Initial Configuration | Total energy (eV) | Magnetic moment ($\mu_B$) |
|---|---|---|---|
| 1 | Cd↑ Se ↓ | 0.00000 | 0.0000 |
| 2 | Cd↑ Se ↑ | 0.00130 | 0.2933 |

The first or the antiferromagnetic spin configuration did not result in any dangling bonds that are not dimerized as shown in Figure 2a) below. Therefore, no magnetic moment was obtained. On the other hand, configuration 2 with initial ferromagnetic coupling resulted in dangling bonds that are not dimerized as seen in Figure 2b) – 2d). It is the non-dimerized dangling bonds that result in magnetic moments.

### 4.3. Surface reconstruction of bare QDs

As seen in the Figures 2a) and 2b) below, the core ions are not displaced and retain their tetrahedral symmetry. In contrast, the surface atoms have undergone reconstructions which we categorize as follows:

#### 4.3.1. Buckling of Cd atoms with CN = 3

The red atoms on the (0001) plane in the Figure 2a) are Cd atoms that have moved from their original positions to a planar configuration, losing their tetrahedral symmetry. Such a reconstruction has been observed across all II-VI and III – V semiconductor compound surfaces and has been attributed to rehybridization of the tri-coordinated cations by Duke.[15,16] The electrons are redistributed from their initial $sp^3$ configuration to $sp^2$ configuration.

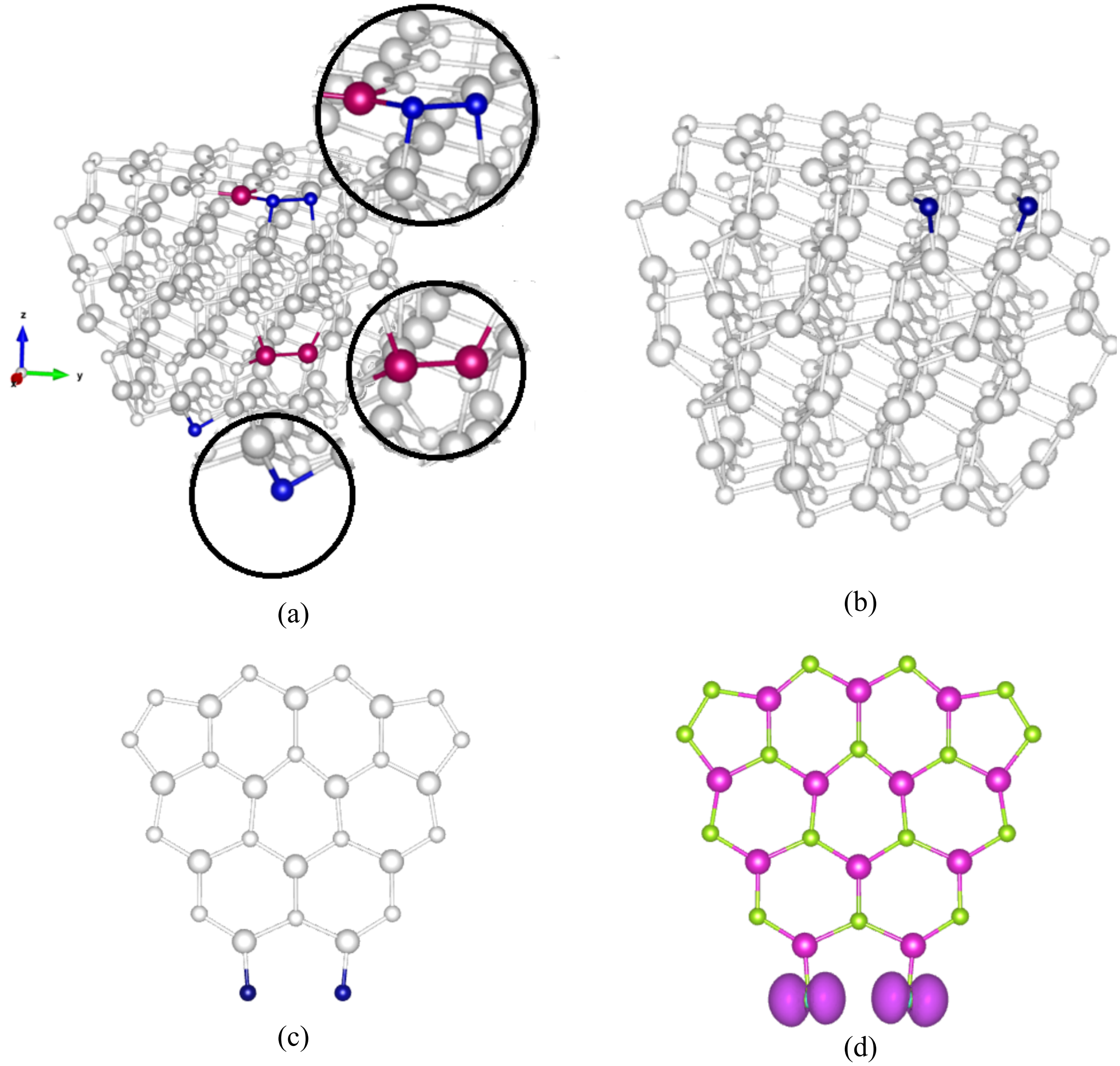

**Figure 2.** Dimerization or non-dimerization of $Cd_{90}Se_{90}$ surface atoms. a) The optimized QD of antiferromagnetic configuration from Table 1. The blue tri-coordinated Se atom has moved out of the QD, the red tri-coordinated Cd atom has moved to a planar configuration, and the bi-coordinated red Cd and blue Se atoms have formed dimers. b) The optimized QD of the ferromagnetic configuration from Table 1. The two un-dimerized Se atoms are in blue. c) Two-dimensional view of the un-dimerized Se atoms. d) Spin density plots, showing the two Se atoms from QD in (c), the purple represents up spin.

4.3.2. Dimerization of the Cd and Se atoms with CN = 2

The blue atoms in Figure 2a) are Se atoms. As seen in the figure, the two Se atoms and the two Cd atoms have formed dimers, thereby losing their tetrahedral configuration, leading to the reconstruction of the surface. Such a reconstruction has also been observed in II-VI and III-V semiconductor compounds and is attributed to the tendency to saturate the valences of the surface atoms by Duke.[15,16]

4.3.3. Outward displacement of Se atoms with CN = 3

The blue Se atoms on the $(000\bar{1})$ plane in Figure 2a) have moved out of the QD. Duke explained the displacements of such anions in semiconductor compounds as the pairing of the electrons in the Se dangling bond due to charge transfer from the tri-coordinated cations.[15,16] The pairing results in a filled s orbital and three unpaired p orbitals. The displacements result in $sp^3$ like pyramid configuration.

The three reconstructions above can explain why the antiferromagnetic coupling of Table 1 result in zero magnetic moments and the ferromagnetic coupling in magnetic moments. The antiferromagnetic configuration result in the optimized structure of Figure 2a) due to reconstruction that displaces Cd and Se atoms with CN=2 and CN=3 that saturates their dangling bonds. As a result, there are no unpaired electrons or dangling bonds that contribute to magnetic moment. The ferromagnetic configuration, on the other hand, has two Se atoms that are not dimerized as seen in 2b), which results in magnetic moment.

4.4. Charge distribution of passivated CdSe QDs

4.4.1. Charge distribution along bonds

Charge density distribution along bonds before or after reconstruction can be figured through charge density difference plots. This is done as follows: the surface of the QD is passivated at all dangling bond sites but one. The uncapped site is chosen at all possible sites of Se and Cd atoms with CN = 3 and CN = 2 on all the facets as listed in Table 2.

Table 2: Sites of uncapped surface atoms with coordination numbers (CNs) and their magnetic moments.

| No | Plane | Atom | CN | Local Magnetic Moment ($\mu_B$) | | | |
|---|---|---|---|---|---|---|---|
| | | | | s | p | d | tot |
| 1 | (0001) | Cd1 | 3 | -0.000 | 0.000 | -0.000 | -0.000 |
| 2 | $(1\bar{1}00)$ | Cd2 | 2 | 0.209 | 0.031 | -0.003 | 0.237 |
| 3 | $(000\bar{1})$ | Se3 | 3 | -0.000 | 0.000 | -0.000 | -0.000 |
| 4 | $(1\bar{1}00)$ | Se4 | 2 | 0.007 | 0.354 | 0.000 | 0.361 |
| 5 | $(11\bar{2}0)$ | Cd5 | 3 | -0.000 | 0.000 | -0.000 | -0.000 |
| 6 | $(1\bar{1}00)$ | Cd6 | 2 | -0.217 | -0.033 | 0.005 | -0.245 |
| 7 | $(11\bar{2}0)$ | Se7 | 3 | -0.000 | 0.000 | -0.000 | -0.000 |
| 8 | $(1\bar{1}00)$ | Se8 | 2 | -0.007 | -0.353 | -0.000 | -0.360 |

The charge density difference ($\delta\rho$) around Cd and Se atoms is then calculated as

$$\delta\rho = \rho_t - \rho_a - \rho_s$$

where $\rho_t$ is the charge density of the QD, $\rho_a$ is the charge density of the isolated atom, $\rho_s$ is charge density of the QD without the chosen atom. If there is an increase or decrease in the charge density along the bonds between the atoms due to motion of charge into or away from the ionic cores, it will manifest in charge density difference plots. A decrease in charge density will imply transfer of charge to a different site such as a dangling bond to pair with an unpaired electron.

In Figure 3, we present first the optimized structures associated with the charge density difference plots.

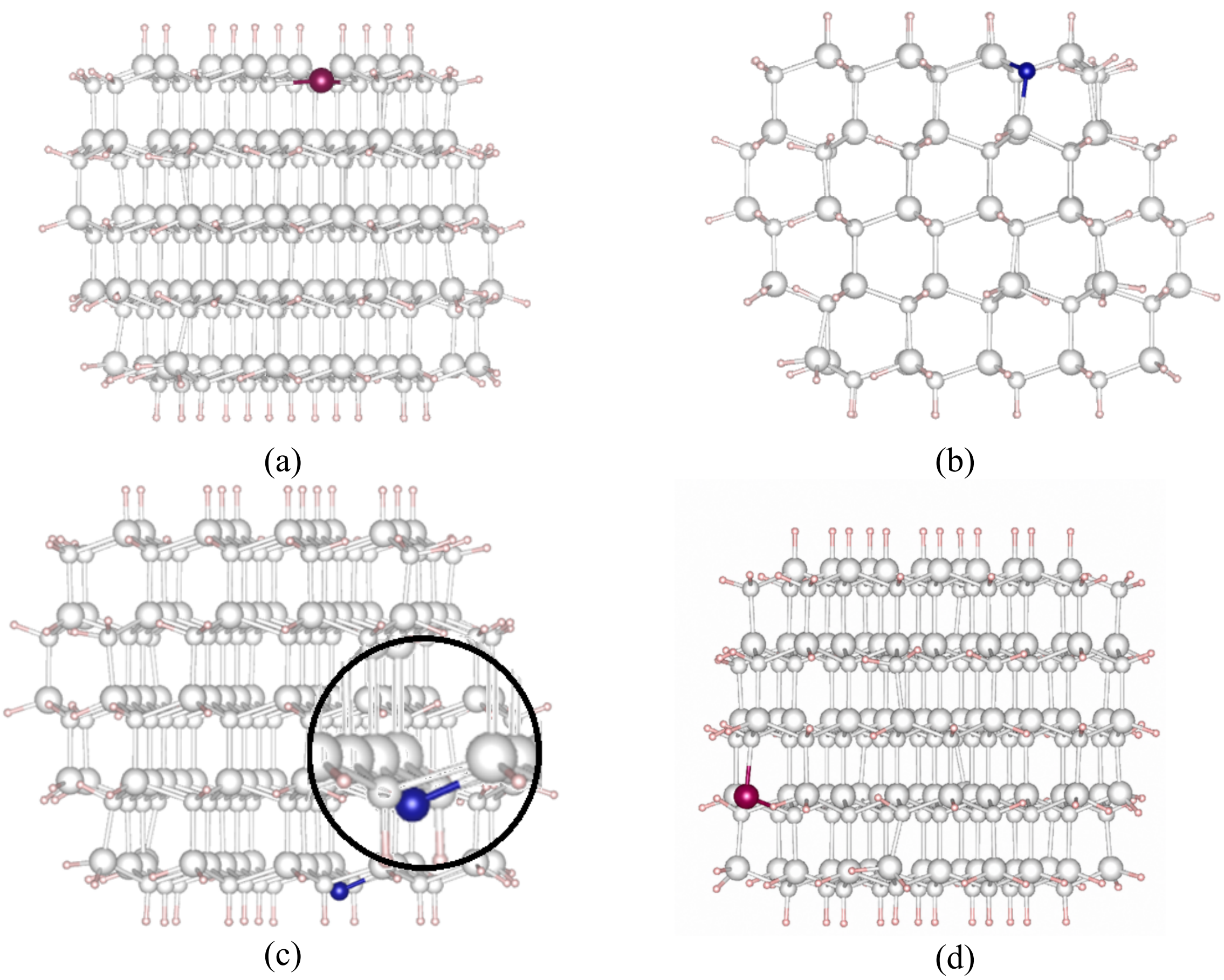

**Figure 3.** The optimized structures of Table 2, all the atoms have the color white except for the uncapped atoms: blue Se and dark brown Cd, (a) tri-coordinated Cd1 (b) bi-coordinated Se4 (c) tri-coordinated Se3 (d) bi-coordinated Cd6.

The tri-coordinated Cd atoms on all planes have rehybridized to a planar configuration as seen in Figure 3a) and the tri-coordinated Se atoms on the planes have moved out of the QD, forming the pyramid-like structure shown in Figure 3c), as in Figure 2b). The bi-coordinated Cd in Figure 3d) and Se atoms in Figure 3b) are not dimerized.

4.4.2. Charge density difference plots at Cd sites

The core Cd atom with CN = 4 is shown in Figure 4c) for comparison. The green color around the Cd atom shows a reduction in the charge density compared to that of an isolated Cd atom. In the case of the tri-coordinated Cd atom shown in Figure 4a), there is an increase in the red color along the bonds between Cd and neighboring Se compared to the Cd atom with a CN = 4 in Figure 4c);

this represents an increase in the charge density along the bonds. The charge density on the other atoms is unchanged. Therefore, we conclude that there is no displacement of electron to another atom or site.

The planar configuration of the Cd atom in Figure 2a) above, with $sp^2$ hybridization, came about without change of charge distribution on neighboring atoms. Therefore, this configuration can be possible only if the two electrons that were being shared across the four $sp^3$ orbitals are now shared across three $sp^2$ orbitals. Thus, an increase in charge density along the bonds and a decrease in bond length from 2.63250 Å to 2.56420 Å, with a stronger covalent bond. This means there are no dangling bonds in the case of tri-coordinated Cd atoms.

Figure 4b) shows a decrease in the red color in the Cd-Se bond compared to Figure 4c), representing a decrease in the charge density and a minimal increase in the bond length by 0.05494 Å. The charge density on the other atoms is unchanged. Therefore, there is no displacement of an electron to another atom. From the two charge distributions of 4a) and 4b) and the nonzero magnetic moment of Cd atom with CN = 2 in Table 2, we conclude the existence of a dangling bond.

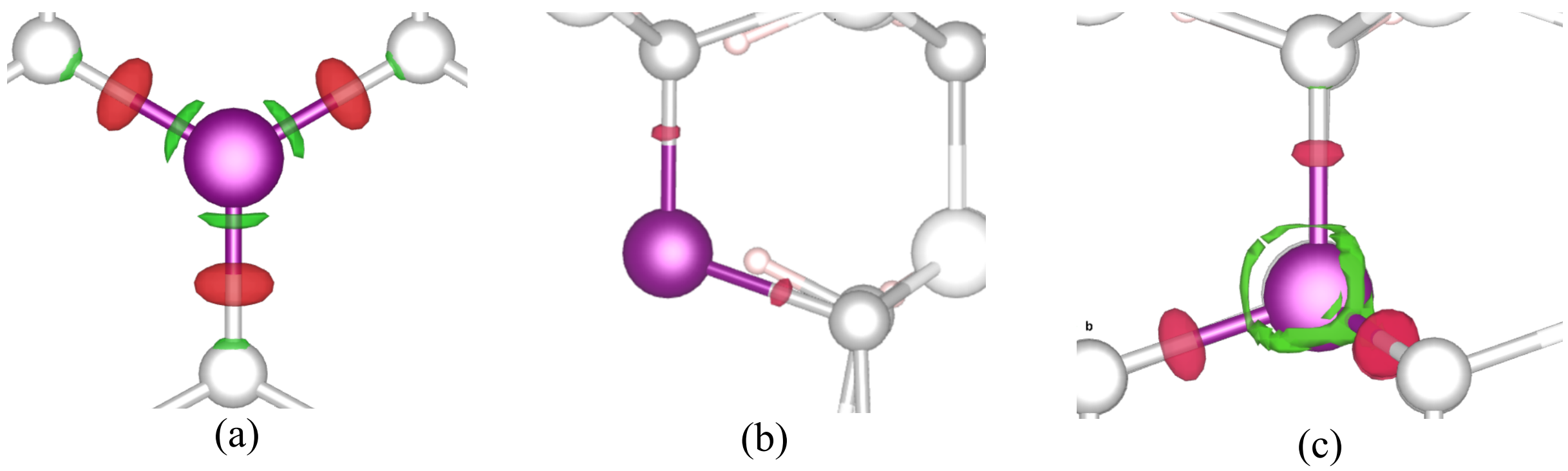

**Figure 4.** Charge density difference ($\delta\rho = \rho_t - \rho_a - \rho_s$) around a Cd surface atom in purple with coordination number of (a) three, (b) two, and (c) a core Cd atom with a coordination number four, all plotted with a value of 0.0075 e/Å$^3$. Red color denotes increase and green denotes a decrease in charge density.

4.4.3. Charge density difference plots at Se sites

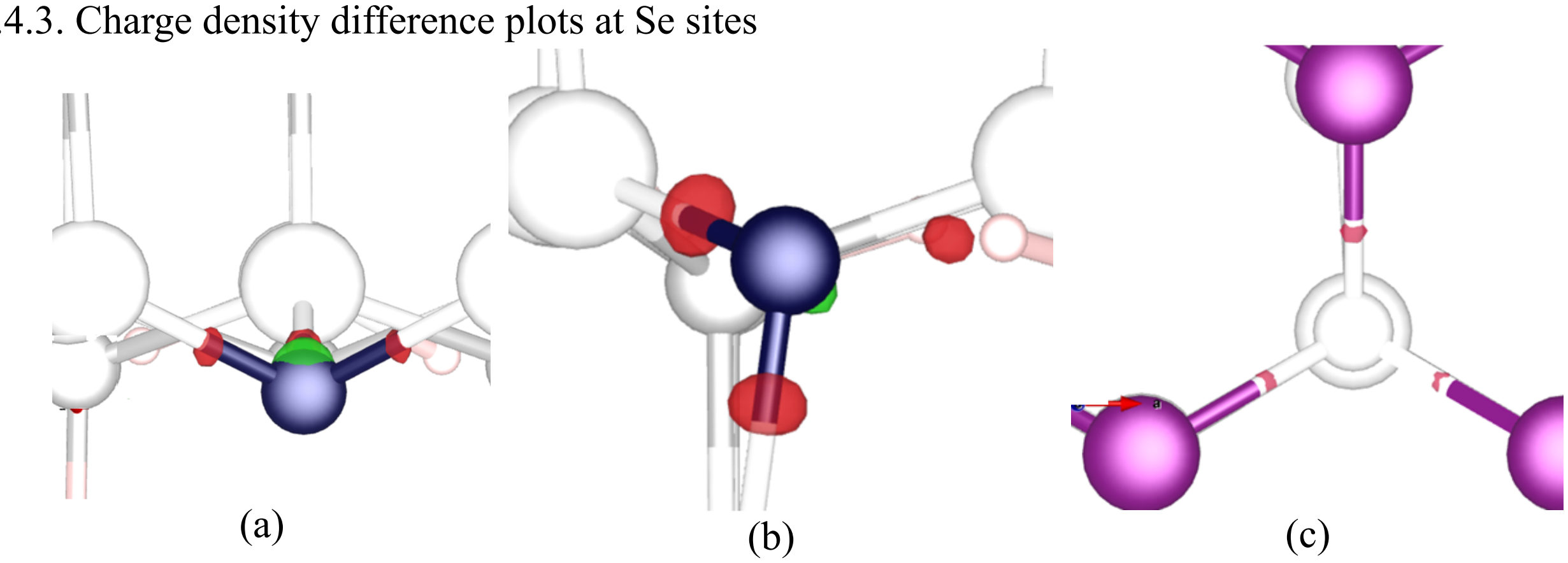

**Figure 5:** Charge density difference ($\delta\rho = \rho_t - \rho_a - \rho_s$) around a Se surface atom in blue with coordination number of (a) three, (b) two, and (c) a core Se atom in white with a coordination number four and plotted with a value of 0.007 e/Å$^3$. Red color denotes increase and green denotes a decrease in charge density.

The Charge density shown in red around the tri-coordinated Se atom in Figure 5(a) shows a slight increase compared to that in Figure 5(c). There is also a more noticeable decrease shown in green, near the tri-coordinated Se atom. The paring of electrons can explain this change in charge density due to 4s and the 4p orbitals hybridizing to form a sp$^3$ bond with the bond length of Cd-Se increasing to 2.7509 Å. The pairing also explains the zero local and net magnetic moment seen in Table 2. Thus, we conclude there are no dangling bonds in the tri-coordinated Se atom.

The Se atom with CN = 2 in Figure 5(b) shows an increase in the charge density between the Cd-Se bonds, shown in red. If each of the sp$^3$ orbital of Se with CN = 4 has a charge contribution of 1.5e from Se and 0.5e charge from Cd to form a bond, then that shared charge of the electron is no longer shared by the dangling bond but by the other remaining Cd-Se bonds leading to an increase in the charge density between them. This arrangement leaves an unpaired electron in the dangling bond, which results in a magnetic moment seen in Table 2.

4.5. Spin density of passivated CdSe QDs

The nature of the magnetic coupling can be figured from spin density plots, the probability of finding polarized spin around the atom. In Figure 6, spin density plots of uncapped sites, from Table 2, with magnetic moments from the bi-coordinated $(1\bar{1}00)$ plane are shown.

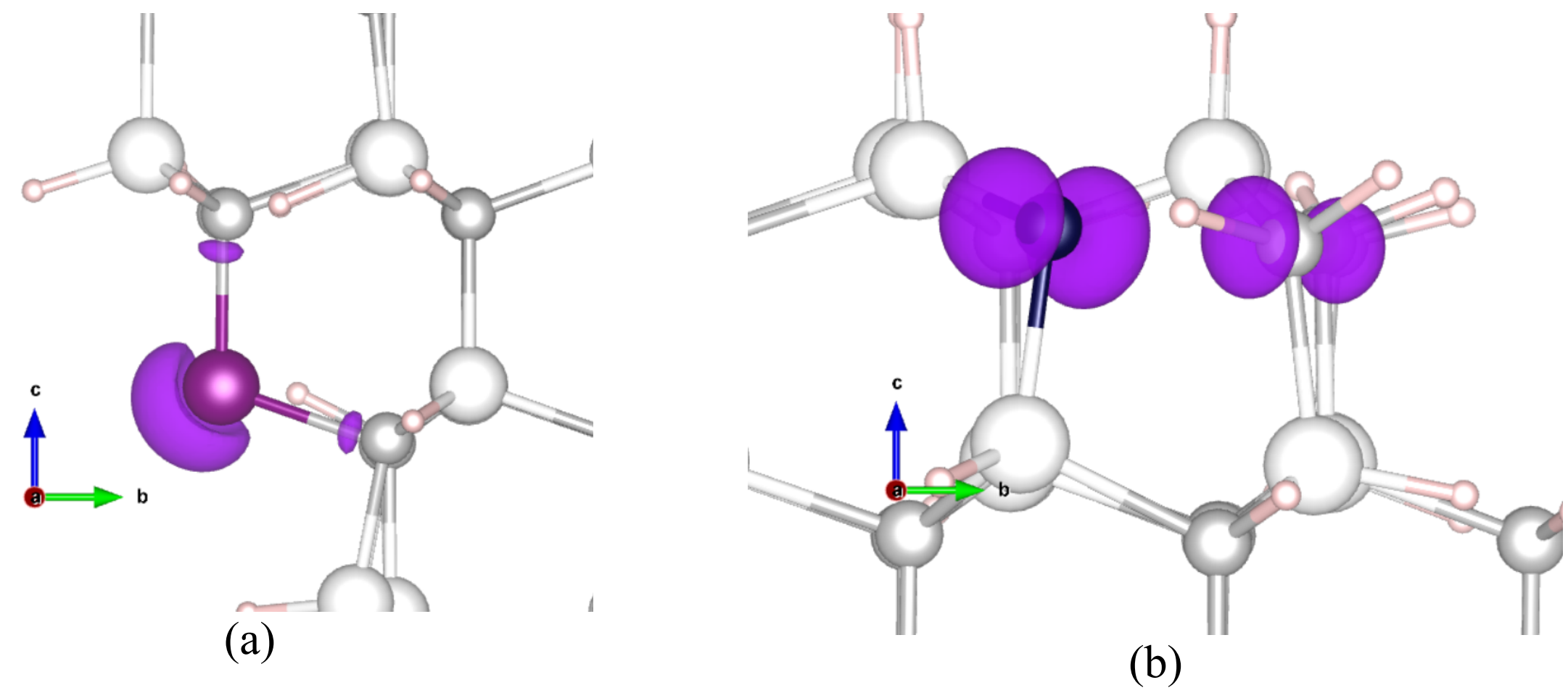

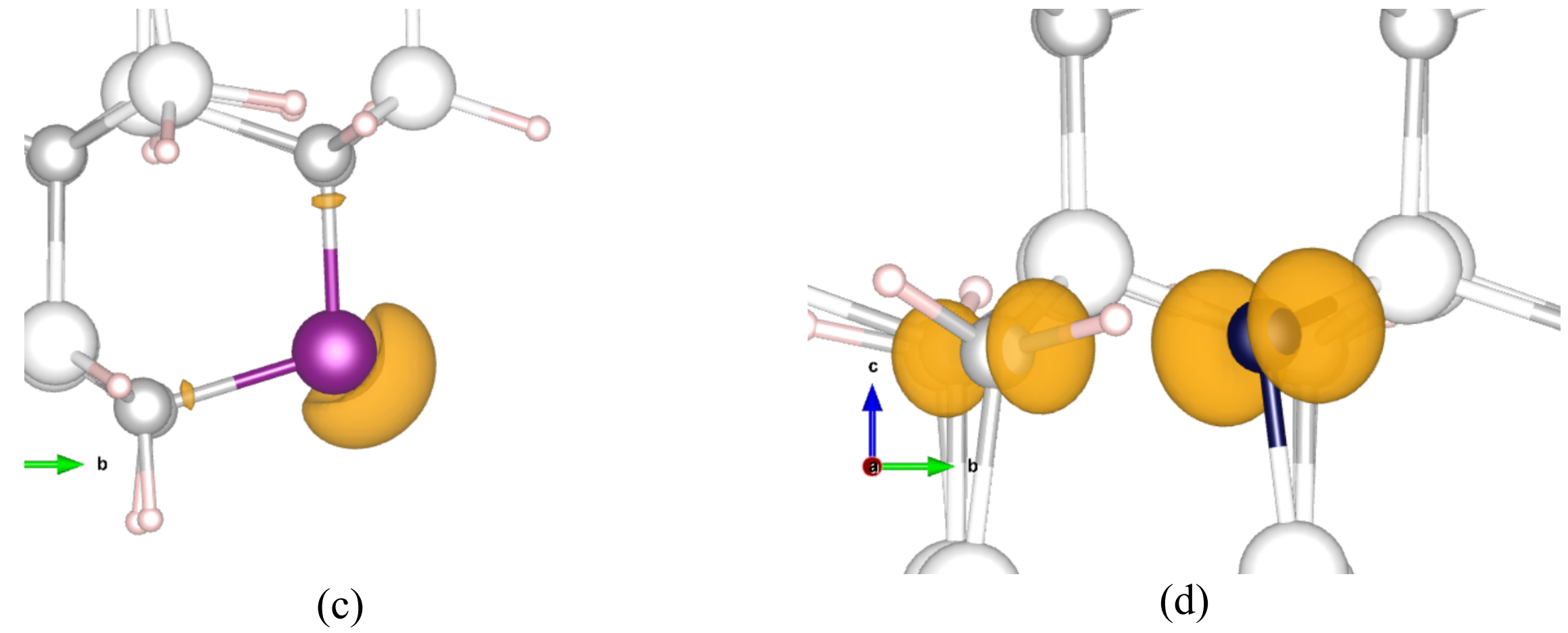

**Figure 6:** The spin density plots for the optimized structures with a nonzero magnetization from Table 2 (a) Cd2, (b) Se4, (c) Cd6, and (d) Se8 plotted with a value of 0.007 e/Ȧ$^3$. Purple color denotes spin up and orange denotes spin down.

Figures 6a and 6c correspond respectively to ferromagnetic coupling between Cd that is uncapped and its neighboring Se, which is quite weak based on relative magnitudes of the spin density plots at the two sites. The spin density at uncapped Cd sites due to the dangling bonds is much larger than the induced spin at the capped Se sites. On the other hand, Figures 6b and 6d correspond to strong ferromagnetic coupling between the uncapped Se site and a neighboring capped Se site; the spin densities at both Se atoms are almost of the same magnitude. The spin – spin correlation between the two Se sites is stronger than that between the Cd and Se sites since Se contributes 1.5e per bond in contrast to Cd that does 0.2e per bond, hence a lager induced spin polarization effect at neighboring Se. Since the dangling bonds in synthesized QDs are attributed mainly to Se atoms, as stated above, the likelihood of ferromagnetic coupling is considerable. The spin orientations in 6a) and 6b) are different from 6c) and 6d) due to the normal to spherical surfaces pointing outward being different in different directions.

4.6. Energy levels and surface states in the band gap of passivated QD

The energy eigenvalues for the configurations in Table 2 are plotted in Figure 7 for fully passivated and partially passivated QDs with one uncapped tri-coordinated and two uncapped bi-coordinated Cd and Se atoms shown in Figures 3, 4, and 5 above. The bandgap is underestimated in DFT calculations, but here we are not interested in the size of the gap, rather in the presence or absence of surface states in the gap. Our computed bandgap of the bulk is 0.494 eV and the HOMO-LUMO gap for the 2 nm CdSe QD is 1.873 eV. This increase in gap is consistent with the quantum confinement effect.

The tri-coordinated Cd and Se atoms do not introduce any states within the gap, while the bi-coordinated Cd atom introduces a state below the LUMO, and the bi-coordinated Se atom introduces a state above the HOMO. The energy levels introduced by the bi-coordinated atoms show the partially ionic nature of the CdSe QD, with anionic Se atom of -2e charge and cationic Cd of +2e charge. When the bi-coordinated Cd atom is uncapped, it acts sort of as a donor with a surface state appearing close to the conduction band edge. The uncapping of the bi-coordinated Cd could involve one or two bonds resulting in the two energy levels seen in the gap that could accommodate possible combinations of spin – polarized surface states. The possibility of such polarized spin states is attractive for spintronic application.

In the case of the tri-coordinated Se atoms, no energy levels are induced within the gap. When the bi-coordinated Se atom is uncapped, spin – polarized surface states are introduced close to the valence band. The surface states created by the atoms with coordination numbers less than four in case of II-VI, and III-V bulk structures were shown[17] and confirmed.[15,16,18] The creation of these surface states indicates the presence of unpaired electrons. The absence of the surface states indicates that no bonds are broken because of the ionic nature of semiconductor QDs as explained, for instance, in the case of ZnO.[19]

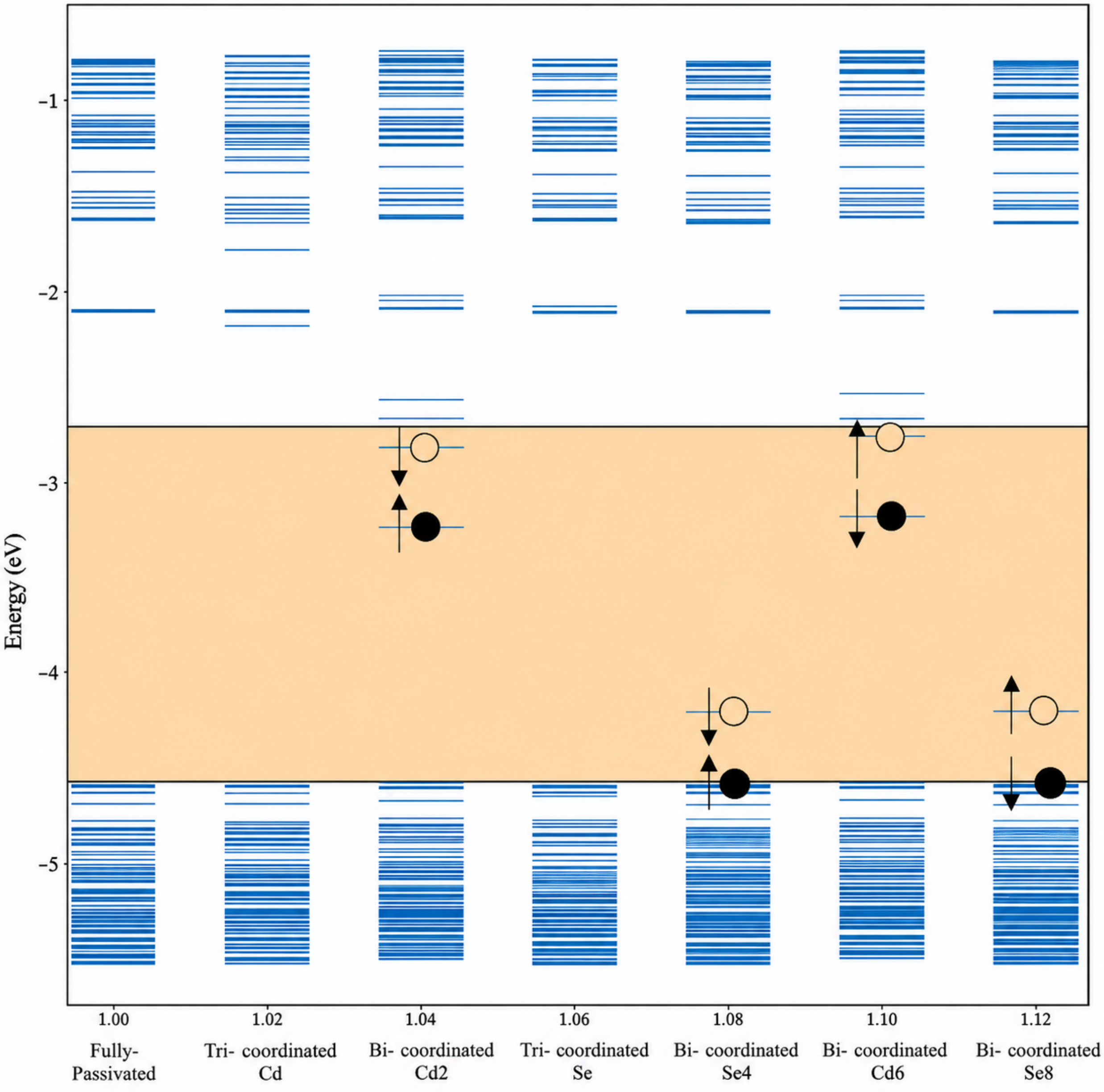

**Figure 7:** Energy level diagrams for fully passivated, partially passivated with one uncapped tri-coordinated Cd atom, bi-coordinated Cd atom, tri-coordinated Se atom, bi-coordinated Se atom. Filled circles show states occupied by electrons, empty circles show states unoccupied by electrons. Spin up states of surface states are indicated by up arrows and the spin down states by down arrows.

### 4.7. Concentration of dangling bonds

Based on the results above, an uncapped tri-coordinated atom will not form a dangling bond and will not contribute to the surface magnetism of the QD. The contribution comes from the bi-coordinated atoms which have not formed dimers or saturated their dangling bonds by making new bonds with neighboring atoms. The bi-coordinated atom forms a dimer if the neighbor is a bi-coordinated atom. If the neighboring atom is not a bi-coordinated atom or is passivated, then the bi-coordinated atom contributes a dangling bond. The symmetry of the quasi-cylindrical structure, seen in Figure 2a), ensures that almost all bi-coordinated atoms have another bi-coordinated atom as their nearest neighbor. Therefore, it is highly unlikely to find more than one or two bi-coordinated Se atoms on the surface that are not saturated, forming a dangling bond. Based on this picture, we estimate the concentration of the dangling bonds.

CdSe QD of 2.8 nm by Neeleshwar et al. has approximately 500 atoms. We carved a CdSe QD out of bulk that is approximately 2.8 nm. It has 438 atoms consisting of around 25 to 28 surface Se atoms with CN = 2 and around 54 to 58 surface Se atoms with CN = 3. This number of uncapped bi-coordinated Se atoms would give a concentration of around 65000 ppm. This high concentration contrary to experimental result suggests that almost all bi-coordinated Se atoms dimerize except one or two. With one or two dangling bonds, the concentration decreases to 2000 to 4000 ppm. This result agrees with the experimental results by Neeleshawar et al of ~2000 ppm for 2.8 nm diameter. The QDs with a diameter of 4.1 nm and 5.6 nm have 1200 and 2200 atoms respectively. With one or two dangling bonds, we get concentrations of 800 ppm for the 4.1 nm and 400 ppm for 5.6 nm, which is close to the experimental results of 1000 ppm and 300 ppm respectively. Thus, our results for the concentration of dangling bonds based on limited number of dangling bonds agree with those by Neeleshwar et.al.

The NMR and XPS studies of CdSe QD topped with TOPO ligands have shown that most Se surface sites are unpassivated.[20] Meulenberg et al. claim that a large concentration of dangling bonds since most Se surface sites are unpassivated. Such an assumption leads to concentrations much higher than those by Neelashwar et al. Although our results for concentration approximately agrees with those of Neelashwar et al, we cannot prove their claim of paramagnetism. Our results, however, show there are dangling bonds with magnetic moments on facets with bi-coordinated atoms, but our colinear calculations cannot determine whether the magnetic coupling is paramagnetic. Non-colinear magnetic calculations are required to determine paramagnetic coupling, which is beyond the scope of this work due to computational cost. However, our colinear, spin – polarized calculations show that ferromagnetic coupling is possible.

### 4.8. Ligands and surface magnetism

The experimental results of X-ray magnetic circular dichroism (XMCD) and X-ray absorption spectroscopy (XAS) by Meulenberg et.al show that CdSe QD doped with TOPO show vacant 4d levels, consistent with a magnetic moment of ~0.01 $\mu_B$ per Cd atom.[21] These results are attributed to the charge transfer from the 4d orbitals of surface Cd atoms to the empty π* orbital of the P═O bond in the ligands. The resulting magnetic ordering was found to be paramagnetic. To test this transfer of charge from the 4d orbital of Cd to the P═O bond in the ligands, we attached a ligand analog $OPC_3H_9$ to a smaller sized QD $Cd_{14}Se_{14}$, as shown in Figure 8, as it is much more costly to compute a larger QD with such a ligand attached.

The ligand analog is attached at different Cd sites shown in Figure 8, while the remaining surface atoms are passivated with pseudo-hydrogens. Magnetic moment calculations resulted in no magnetic moment for all the configurations (including Cd2, Cd3, and Cd4 with ligand analog attached) in Figure 8. Thus, there is no magnetic moment on the surface Cd atoms. This may be due to the very small magnetic moment of ~0.01 $\mu_B$ per Cd atom.[21]

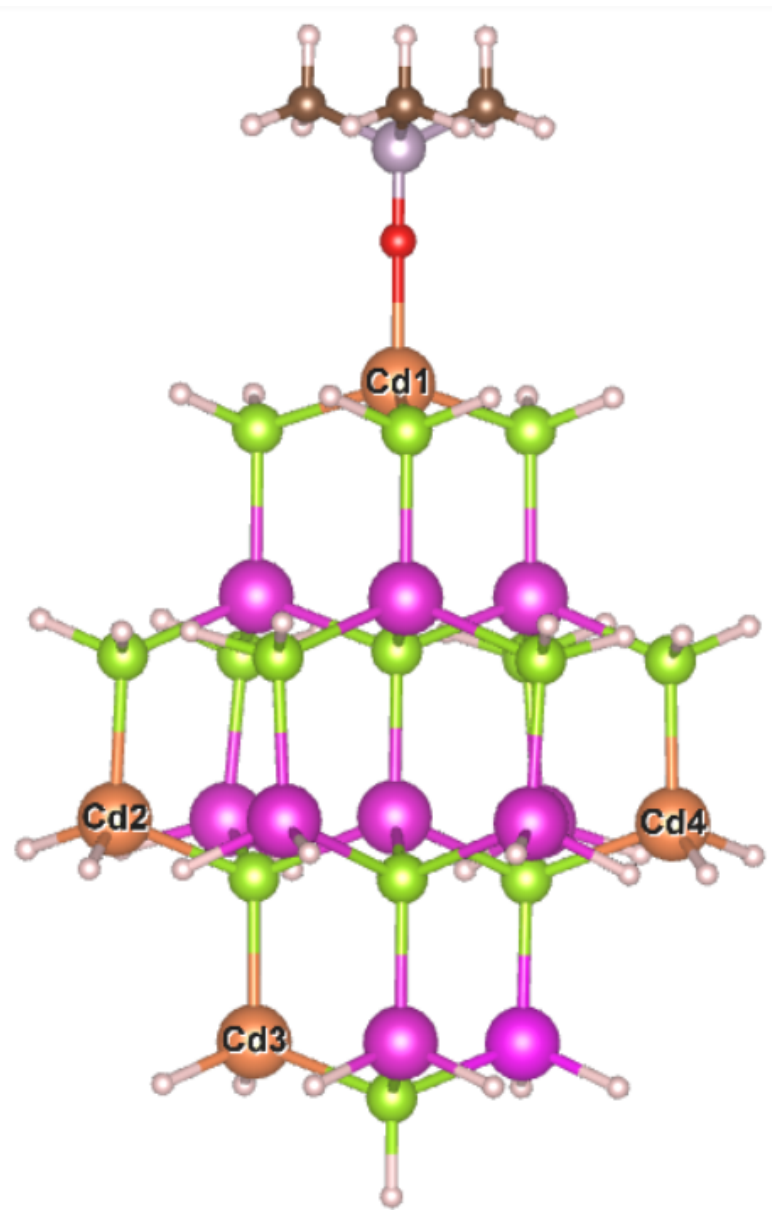

**Figure 8.** The optimized $Cd_{14}Se_{14}$ QD capped with a simple capping group $OPC_3H_9$, at Cd1. The other locations of capping are Cd2, Cd3, and Cd4.

In order to analyze the electronic structure of the Cd capped with the ligand analog, we compared the projected density of states (PDOS) of the d orbitals of the core Cd and the capped Cd as shown in Figure 9. The DOS of the Cd atom capped with the ligand in blue has broadened compared to that of the core Cd atom that is localized, as shown in the DOS plots in Figure 9. The peaks in blue can be seen around the red peak of the core atom. Similar broadening of the DOS is observed among the other Cd sites when capped with $OPC_3H_9$. The results show that there are variations in the local DOS of the d orbital which agree with the experimental results by Meulenberg et.al and DFT calculations.[22] We argue that the overall electronic structure, except the broadening, is similar between the capped Cd and core Cd and the transfer of charge from Cd that could have been evident in change in the structure of PDOS is not seen.

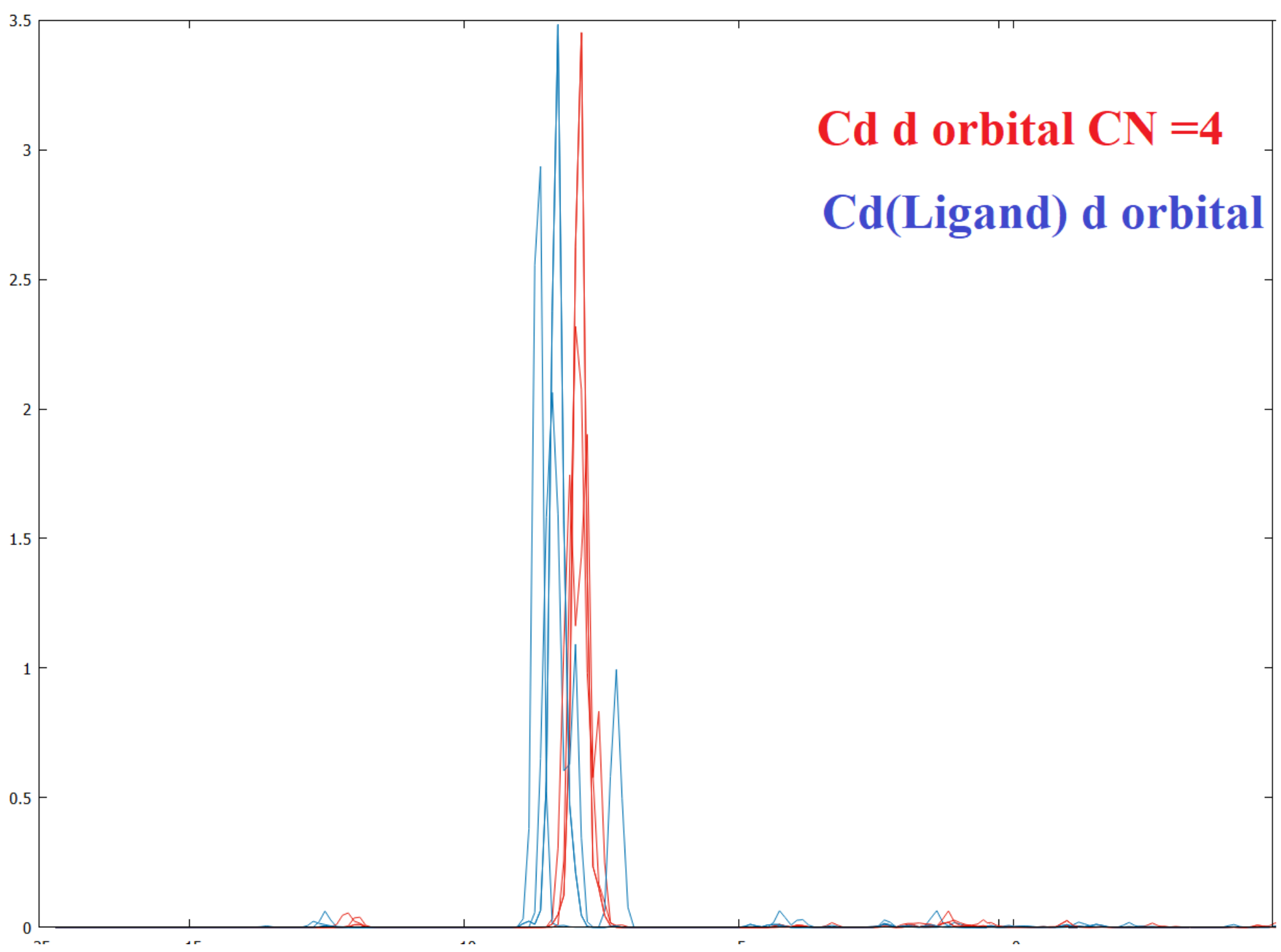

**Figure 9.** Projected Density of states (PDOS) of the d orbital of Cd from the core plotted in red and the d orbital of Cd capped with a ligand plotted in blue.

## 5. Conclusion

Our results show that the induced magnetism observed on undoped CdSe QD is attributed to dangling bonds. The contribution to the induced magnetism is from the surface atoms that have CN = 2. Although CdSe QD capped with TOPO ligands has a large number of uncapped Se surface atoms, the Se atoms with CN = 3 relax, as a result have no magnetic moments. In addition, the Se atoms with CN = 2 dimerize reducing the population of dangling bonds with magnetic moments. The remaining dangling bonds from bi-coordinated Se sites can couple ferromagnetically to neighboring Se atoms by inducing spin polarization according to our colinear spin-polarized calculation. Our results indicate that the relaxation of the hybridized orbitals takes place by redistribution of the charge along the bonds of the same atom or site, and that there is no transfer of charge from the dangling bonds to neighboring sites. The concentration of dangling bonds we determined agrees with the experimental results by Neelashwar et.al. Our calculations on possibility of the ligands contributing to the surface magnetism show that even though there are variations on the local DOS of the surface Cd atoms that are capped by $OPC_3H_9$ group, their magnetic moment is zero. Therefore, based on our colinear spin calculations, we attribute the surface induced magnetism to dangling bonds for CdSe QD capped with TOPO not to charge transfer between Cd and TOPO.

## Acknowledgement

A.A. and M.D.M were supported by the National Science Foundation Grant No. DMR-2013854. D.S. is partially supported by same grant. M.D.M. acknowledges computer time allocation (No. TG-DMR100055 and TG-PHY250389) for Stampede3 at Texas Advanced Computing Center from the Advanced Cyberinfrastructure Coordination Ecosystem: Services & Support (ACCESS) program, which is supported by National Science Foundation grants #2138259, #2138286, #2138307, #2137603, and #2138296.

## Conflict of Interest

The authors have no conflict to disclose.

## Data Availability

The data that support the findings of this study are available from the corresponding author upon reasonable request.